\documentclass[5p]{elsarticle}

\usepackage{amssymb}

\journal{Nuclear Physics A}

\begin{document}

\begin{frontmatter}



\title{Complex Systems: \\
From Nuclear Physics to Financial Markets}


\author[a,b]{J.~Speth}
\ead{j.speth@fz-juelich.de}
\author[b,c]{S.~Dro\.zd\.z}
\author[a]{F.~Gr\"ummer}

\address[a]{Institut f{\"u}r Kernphysik, Forschungszentrum J\"ulich, D-52425 J{\"u}lich, Germany}
\address[b]{Institute of Nuclear Physics , PL-31-342 Krak\'ow, Poland}
\address[c]{Institute of Physics, University of Rzesz\'ow, PL-35-310 Rzesz\'ow, Poland}

\begin{abstract} We compare correlations and coherent structures in nuclei and financial markets. In the nuclear physics part we review giant resonances which can be interpreted as a coherent structure embedded in chaos. With similar methods we investigate the financial empirical correlation matrix of the DAX and Dow Jones. We will show, that if the time-zone delay is properly accounted for, the two distinct markets largely merge into one. This is reflected by the largest eigenvalue that develops a gap relative to the remaining,
chaotic eigenvalues. By extending investigations of the specific character of financial collectivity
we also discuss the criticality-analog phenomenon of the financial log-periodicity and show specific examples.

\end{abstract}

\begin{keyword} Financial Physics, Correlations, Cross-correlations, log-periodicity, collectivity in nuclei



\end{keyword}

\end{frontmatter}


\section{Introduction}
\label{1}
By its very nature the notion of complexity, which is central to the contemporary physics, still lacks a precise definition. The recent concept of the scale free complex networks~\cite{Barabasi} offers one promising direction to formalize complexity though many 
related issues still remain open. In qualitative terms the notion of complexity refers to the diversity of forms, to the emergence of coherent patterns out of randomness and also to the ability of frequent switching among such patterns. This normally involves many components, many different space and time scales and thus phenomena like chaos and noise. Complexity includes of course also collectivity and criticality \cite{Stanley99}. In fact, due to all those elements, it seems most appropriate to search for real complexity just at the interface of chaos and collectivity \cite{Bak}. Indeed, these two seemingly contradictory phenomena have to go parallel, as both are connected with the existence of many degrees of freedom and strong, often random, interactions among them. The random matrix theory (RMT) \cite{Mehta} - a concept originating from nuclear physics considerations~\cite{Wigner} - 
proves very fruitful in approaching complex quantum systems and in addressing the question of how classical chaos manifests itself on the quantum level. Chaos is essentially a generic property of complex systems such as e.g. atomic nuclei and finds evidence in a broad applicability of RMT to describe level fluctuations \cite{Weidi}.

 Based on two natural complex dynamical systems, the strongly interacting nuclear many-body system and the financial markets we present arguments in favor of the statement above that complexity is a phenomenon that involves both, collectivity and chaos. We also show that in the financial markets collectivity in addition carries signatures of criticality.
 
\section{The Nuclear Many-Body System}

\begin{figure}
\begin{center}
{\includegraphics[height=10cm]{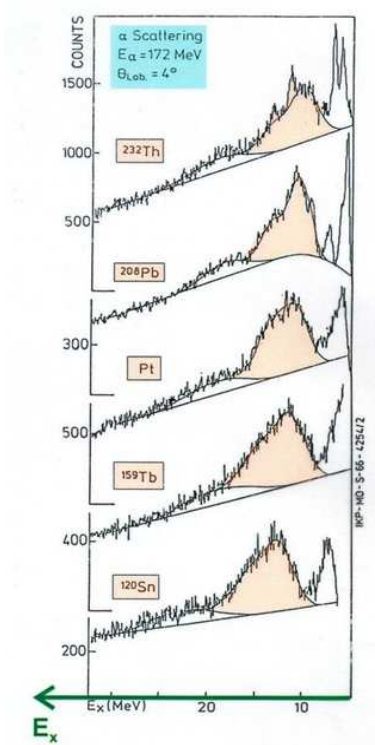}}
\caption{\label{fig1} Giant quadrupole resonances in atomic nuclei. }
\end{center}
\end{figure}

\begin{figure}
\begin{center}
{\includegraphics[height=8cm,width=6cm]{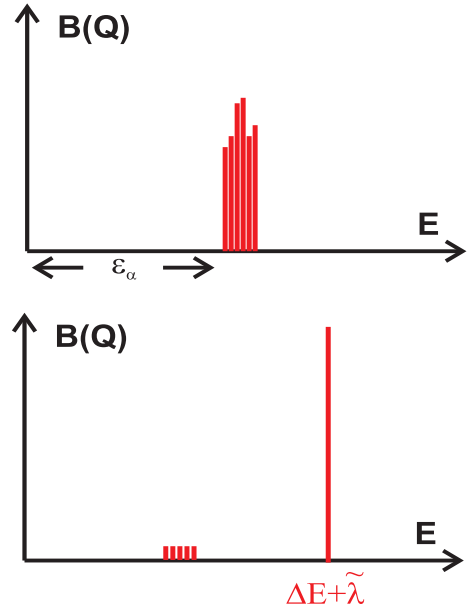}}
\caption{\label{fig1a} RPA solution in the schematic Brown-Bolsterli model: In the upper part the degenerate 1p1h states are shown, in the lower part the corresponding RPA solution for a repulsive ph-interaction. }
\end{center}
\end{figure}

The structure of nuclei is an excellent and well known example for collectivity embedded in chaos \cite{Stan95,Stan98}. On one hand the single particle structure in odd mass nuclei, the rotational spectra in deformed nuclei, the low-lying collective $2^+$ and $3^-$ states in spherical nuclei as well as the high-lying giant resonances are indications of a partial regular behavior in nuclei. The absolute number of such states in each nucleus, however, is small compared with the total number of levels. For that reason statistical methods have been developed and used to analyze the remaining huge number of individual levels \cite{Weidi} that indicate chaotic behavior. The best example for collectivity embedded in chaos are giant resonances. These collective structures exist in all medium and heavy mass nuclei. In Fig.~\ref{fig1} the electric giant quadrupole resonance in several nuclei is shown. Collectivity means a cooperation, and thus the coupling between the different degrees of freedom in order to generate a coherent signal in response to an external perturbation. In shell model approaches collective states may be generated within the random phase approximation (RPA) in a 1p1h configuration space \cite{Speth77}. For a quantitative description of the width of giant resonances, however, it is necessary to couple to the 2p2h space, which is crucial for an appropriate description of their decay properties \cite{Stan90}.
Interestingly, due to a sufficiently large density of states relative to the strength of the residual interaction, local level fluctuations characteristic for a Gaussian orthogonal ensemble (GOE) appear already in the space of 2p2h states \cite{Stan95,Stan98}. In other words: The collective 1p1h part of the giant resonances is embedded in the chaotic 2p2h space. It is important to point out that the coherent structure in the  giant resonance case is due to the long-range correlations in nuclei. 
The coherent solution of the RPA in nuclei has its analogue in the stock market as we will see in the following section. The mechanism that creates such coherent solutions can be best understood in a schematic model developed by Brown and Bolsterli \cite{Gerry}. Here one assumes that all 1p1h states are energetically degenerate as indicated in the upper part of Fig.~\ref{fig1a}. As interaction one uses a separable multipole-multipole interaction which allows to solve the RPA equation analytically. The result is shown in the lower part of Fig.~\ref{fig1a}. Only one state is strongly shifted in energy and includes all the transition strength $B(Q)$. All the other states remain at their original energies and loose nearly all their transition strength. In the present example the interaction is repulsive, if the interaction is chosen attractive the coherent state is shifted to lower energies. In a realistic case the coherent state carries a large fraction of the strength and one is able, if the 2p2h configuration space is properly included, to explain the experimental spectra shown in Fig.~\ref{fig1}. 


\section{Correlations in the Stock Market}

\begin{figure}
\begin{center}
{\includegraphics[width=9cm]{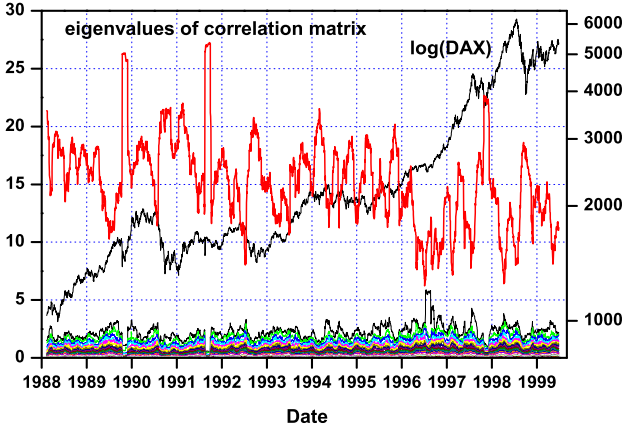}}
\caption{\label{fig1b} Time-dependence of the eigenspectrum of the DAX correlation matrix during the years 1988-1999. The left abscissa is the total portfolio variance $\sigma_P$ defined in Eq.~(\ref{eq:1c}) and the right is the value of the DAX. The thick (red) line represents the most collective solution and the thin line from the lower left corner to the upper right corner indicates the DAX.}
\end{center}
\end{figure}

It is interesting to investigate correlations in financial markets with theoretical tools similar to that used in nuclear physics. We investigated correlations within the Deutsche Aktienindex (DAX) and the Dow Jones industrial average (DJIA) \cite{Stan00} and cross correlations between the two indices \cite{Stan01}. For this investigation the empirical correlation matrix $\bf{C}$ of the 30 companies (in both indices) was constructed during the period of eleven years. A time window of 30 trading days was considered. The matrix elements $C_{ij}$ of the correlation matrix are defined as:
\begin{eqnarray}\label{eq:1a}
C_{ij}~=~\frac{\left\langle G_i(t)G_j(t)\right\rangle_t~-~\left\langle G_i(t)\right\rangle_t\left\langle G_i(t)\right\rangle_t}{\sigma(G_i)\sigma(G_j)}.
\end{eqnarray}
Here $G_i(t)$ is the return for a price $x_i(t)$ of the $i^{th}$ asset at time t defined as:
\begin{eqnarray}\label{eq:1b}
G_i(t)~=~ln(x_i(t+\tau))-ln(x_i(t))
\end{eqnarray}
where $\tau$ is called the time lag and $\sigma(G_i)$ is the volatility of $G_i(t)$. If one diagonalizes the correlation matrix one obtains the result shown in Fig.~\ref{fig1b} which is very similar to the one shown in Fig.~\ref{fig1a}. First of all there exists one linear combination which is completely different from the other 29 solutions. This solution follows very closely the drawups and drawdowns of the DAX but in a significant distinct way. The drawdowns are dominated by one strongly collective eigenstate with a large eigenvalue. Such a state exhausts a large fraction of the total portfolio variance:
\begin{eqnarray}\label{eq:1c}
\sigma_P~=~\sum_{i,j}^{N}p_iC_{ij}p_j
\end{eqnarray}
where $p_i$ expresses a relative amount of capital in the asset $i$. The stronger the correction in the index, the more pronounced is the effect. At the same time, due to the conservation of the trace of $\bf{C}$ the remaining eigenvalues are close to zero. The opposite applies to drawups. Their onset is always accompanied by a sizable redistribution in the eigenvalues towards a more uniform distribution. We may conclude: The increase of the stock market never involves parallel uniform increases of prices of all the companies as it happens during decreases. One obtains exactly the same result for the Dow Jones index.
 
 From daily experience one knows that the European stock market follows very closely the American stock market. The well known fact can be quantified with help of the global $\cal G$ correlation matrix $\bf{C}_{\cal G}$ of the DAX and Dow Jones:
\begin{eqnarray}\label{eq:1d}
\bf{C}_{\cal G}~=~ \left ( \matrix{ {\bf C}_{DAX,DAX} &  {\bf C}_{DAX,DJ} \cr
                 {\bf C}_{DJ,DAX}  &   {\bf C}_{DJ,DJ} } \right).
\end{eqnarray}

\begin{figure}
\begin{center}
{\includegraphics[width=6cm,angle=270]{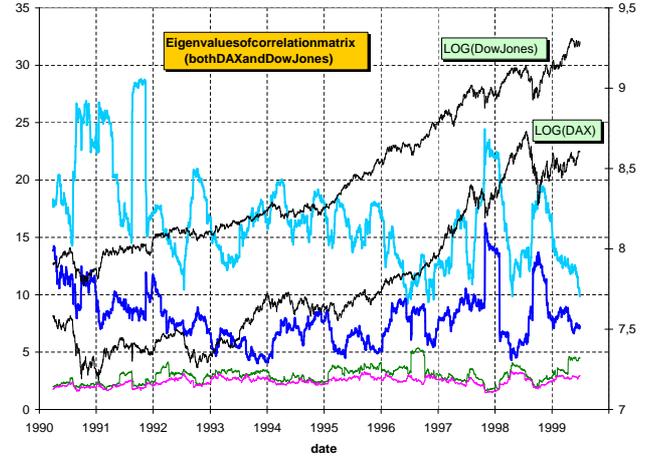}}
\caption{\label{fig1d}~Time-dependence of the eigenspectrum of the global (DAX+DJ) correlation matrix ${\bf C}_{\cal G}$ calculated from the time-series of daily price changes in the interval of T~=~60 past days, during the years 1990-1999. The left abscissa is the total portfolio variance $\sigma_P$ and the right denotes the logarithm of the DAX and DJ indices. The two solutions are essentially the independent results for the DJ (upper curve) and DAX (lower curve).}
\end{center}
\end{figure}
\begin{figure}
\begin{center}
{\includegraphics[width=6cm,angle=270]{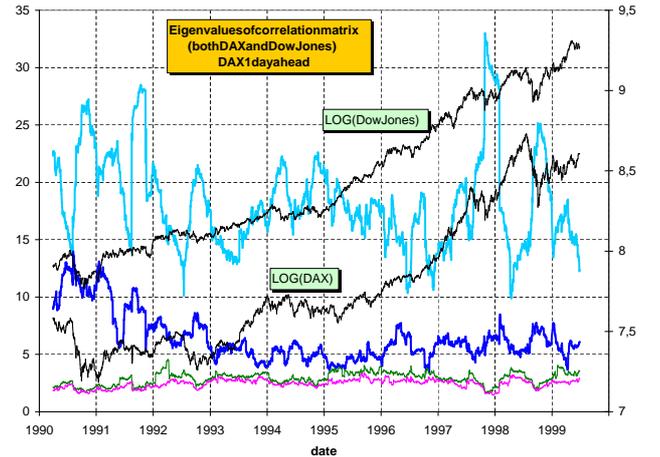}}
\caption{\label{fig1c} Same as in Fig.~\ref{fig1d}  but with the DAX returns shifted one day ahead relative to the DJ. Here one obtains one single collective solution, which includes DJ and DAX.}
\end{center}
\end{figure}
In the present case, where one incorporates all the stocks traded by DAX and by Dow Jones, one has to consider 60 trading days. The total time-interval explored here includes the years 1990-1999. If one considers naively the daily data of the DAX and Dow Jones as done in the calculation which gives the results shown in Fig.~\ref{fig1d}, one obtains two collective solutions which represent the two stock markets as if they were essentially independent. This is more strongly pronounced after 1996, when the stock market started to rise. 

If one takes, however, the DAX data one day in advance relative to the Dow Jones returns, one obtains a significantly different eigenvalue spectrum, as shown in Fig.~\ref{fig1c}. Now one single eigenvalue dominates the dynamics which means that a sort of common market emerges. The explanation is obvious: because of the time difference between New York and Frankfurt the DAX is strongly influenced by the Dow Jones results from the previous day.

\section{Log-periodic Structures in the Financial Markets}

In the previous sections we have demonstrated the coexistence of collective (coherent) structures and of chaos both in nuclei and in the financial markets. We now address a question concerning the more 
specific character of the financial collectivity.   
Like the nuclear many-body problem, also the financial dynamics is a multiscale phenomenon and therefore the question which properties are scale invariant and which are scale characteristic refers to the essence of this phenomenon. There exists strong evidence that at least a part of the financial dynamics - in fact the one that is associated with the largest eigenvalue of the correlation matrix 
- is governed by phenomena analogous to criticality in the statistical physics sense \cite{Sornette}. Criticality implies a scale invariance \cite{Stanley71} of a properly defined function 

\begin{eqnarray}\label{eq:1}
\Phi(\lambda x)~=~\gamma \Phi(x)
\end{eqnarray}

The constant $\gamma$ describes how the properties of the system change under rescaling by the factor $\lambda$. One obvious solution of Eq.~(\ref{eq:1}) is a power-law. The zig-zag character of the financial dynamics, however, attracts attention to a more general solution of Eq.~(\ref{eq:1}) that reads \cite{Jona}
\begin{eqnarray}\label{eq:2}
\Phi(\ x)~=~x^{\alpha}\Pi(ln(x)/ln(\lambda)).
\end{eqnarray}

\begin{figure}
\begin{center}
{\includegraphics[width=8cm]{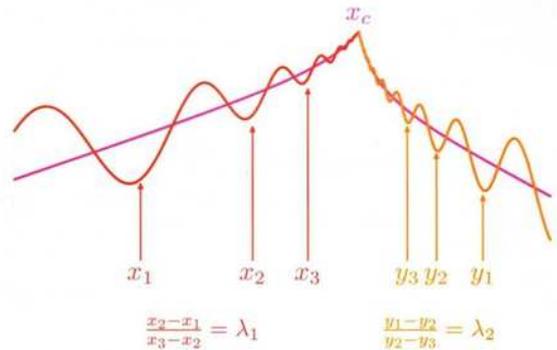}}
\caption{\label{fig2}Log-periodic modulations}
\end{center}
\end{figure}

The first term represents a standard power-law that is characteristic of continuous scale-invariance with the critical exponent $\alpha~=ln(\gamma)/ln(\lambda)$ and $ \Pi(x)$ denotes a periodic function of period one that can be interpreted in terms of discrete scale invariance. The second gives rise to a correction of the conventional scaling that is periodic in $ln(x)$. Imprints of such oscillations can often be identified in the financial dynamics \cite{Sornette96,Feigenbaum,Vandervalle,Stan99}. In the case of financial markets one defines $x~=~\left|T-T_c\right|$ where $T$ is the clock time of the price time series and $x$ represents the distance to the critical point $T_c$. The emerging sequence of spacing between the minima $(x_1,x_2,x_3)$ shown in Fig.~\ref{fig2} form a geometric progression according to the ratio $\lambda_1$ for the accelerating bubble market phase on the left hand side of the reversal point $x_c$ and $\lambda_2$ for the decelerating anti-bubble phase on the right hand side, both defined in Fig.~\ref{fig2}. For the financial markets $\lambda_1=\lambda_2\approx2$ \cite{Stan03,Bartolozzi}. An important related element, for a proper interpretation and handling of the financial patterns, as well as for the consistency of the theory, is that such log-periodic oscillations manifest their action self-similarly through various time scales \cite{Stan99}. This applies to the log-periodic accelerating bubble market phase as well as to the decelerating anti-bubble phase.

A specific form of the periodic function $\Pi(x)$ in Eq.~(\ref{eq:2}) is not provided by any first principles. One reasonable possibility is the first term of its Fourier expansion,

\begin{eqnarray}\label{eq:3}
\Pi(ln(x)/ln(\lambda))~=~A+B\cos(\frac{\omega}{2\pi}ln(x)+\Phi).
\end{eqnarray}
Another approach to log-periodic oscillations follows from a generalization of the conventional random walk, the so called Weierstrass random walk. In the conventional random walk there is a single step length $a$ and one probability $c$. In the Weierstrass random walk one allows steps not only of one single length $a$ but of multiple lengths $b^ja$. These longer jumps happen with a reduced probability $c/M^j$. The two parameters are restricted by $(b>1$,$M>1)$. The conventional random walk gives rise to a Gaussian distribution, whereas the probability distribution in the Weierstrass \cite{Weierstrass} random walk is given by the Weierstrass function, 
\begin{eqnarray}\label{eq:4}
p(k)~=~\frac{M-1}{2M}\sum_{j=o}^{\infty}\frac{1}{M^j}\cos(kb^ja).
\end{eqnarray}

\begin{figure}
\begin{center}
{\includegraphics[width=8cm]{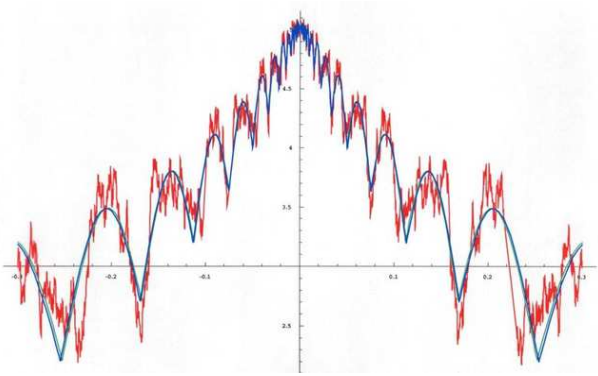}}
\caption{\label{fig3}Computer-simulation of the Weierstrass function and the analysis with the function given in Eq.~(\ref{eq:3}).}
\end{center}
\end{figure}

\begin{figure}
\begin{center}
{\includegraphics[width=10cm]{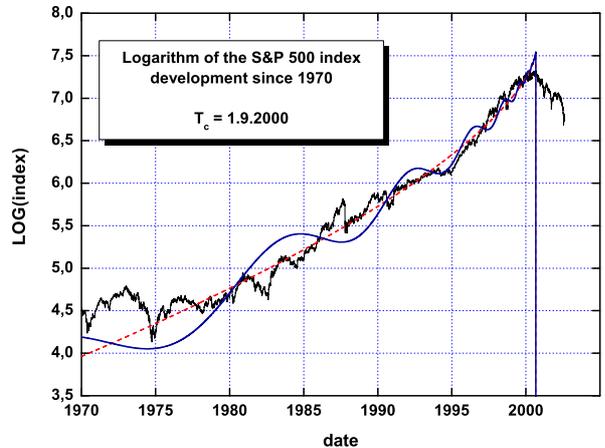}}
\caption{\label{fig4} Logarithm of the Standard \& Poor's over the period 1970-2002 and its log-periodic analysis (solid line). The dashed line represents a power-law.}
\end{center}
\end{figure}

\begin{figure}
\begin{center}
{\includegraphics[width=10cm]{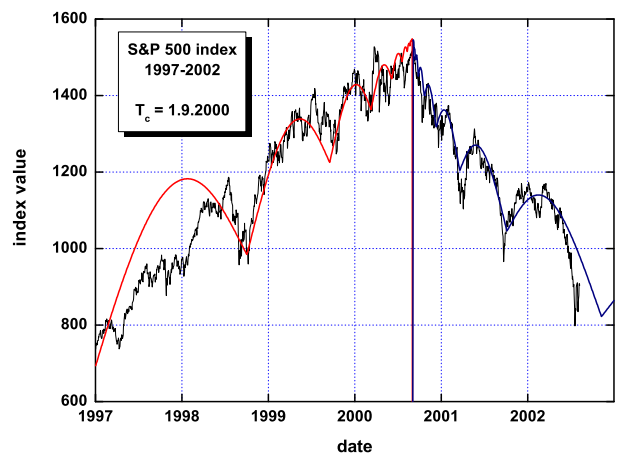}}
\caption{\label{fig5}Standard \& Poor's over the period from 1997 till the end of August 2002 and its log-periodic analysis. The straight solid line serves as a reference illustrating the log-periodically accelerating and decelerating scenarios with a common time of the crash $T_c$=1.9.2000.}
\end{center}
\end{figure}

The Weierstrass random walk possesses a discrete self-similarity as one can see from Fig.~\ref{fig3}. Here a numerical solution of the Weierstrass function is presented that shows the typical log-periodic structure. The numerical solution is analyzed
with the log-periodic ansatz given in Eq.~(\ref{eq:3}). This function reproduces the minima and maxima in an excellent way.
 In the following we present various analyses of the financial markets using simplified parameterizations for the periodic function $\Pi(x)$: we use the cosine as well as the modulus of the cosine to describe the oscillations.
In Fig.~\ref{fig4} we present an analysis of the Standard and Poor's 500 index in the period $1970-2002$. The data start in $1970$ and the two most prominent minima, where we begin with, are the deep correction at the end of 1974 and the famous Black Monday of October $19,1987$. The log-periodic function remains systematically in phase with the following market trend and points to September $1,2000$ as the date of the reversal of the almost 20 years global increasing trend. In this respect the two deep minima, are prominent log-periodic precursors of a more serious crash that started in September 2000. It is clear, bearing in mind the uncertainties connected with the analysis, one could have predicted a dramatic change in the second half of 2000. A closer inspection of the vicinity of this date, obtained by the magnification presented in Fig.~\ref{fig5}, shows two other relevant elements. It clearly indicates, first of all, that the modulus of the cosine in Eq.~(\ref{eq:3}) provides a better representation for the log-periodic modulation in the present case. Secondly, it provides independent evidence that the real date $T_c$ marking the reversal of the upward global trend at the beginning of September 2000, is simultaneously also the date when the decelerating log-periodic oscillations accompanying the declining starts.
This spectacular agreement of the end of the bubble phase with the beginning of the anti-bubble phase is an extra argument in favor of the consistency of the log-periodic scenario. It is also amazing how well the oscillation pattern in Fig. \ref{fig5} agrees with the computer simulation of the Weierstrass function shown in Fig.~\ref{fig3}.
 
\begin{figure}
\begin{center}
{\includegraphics[width=10cm]{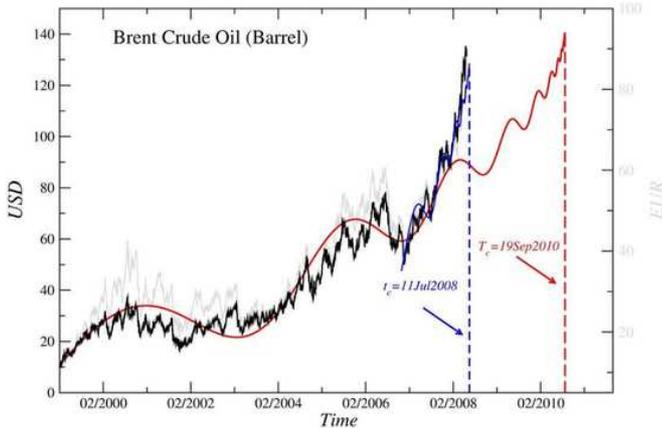}}
\caption{\label{fig6} This figure has been published on the Los Alamos preprint server on June 23,2008 as an insertion (note added) to ref \cite{Stan08}. The explanation of the figure is given in the text.}
\end{center}
\end{figure}

 Other very valuable testing grounds of what we call \emph{the prediction oriented variant of financial log-periodicity} \cite{Stan04} 
are the violent price changes and the related speculative bubbles in the world commodity markets in the past years such as the gold market or the oil market. As an example we present in Fig.~\ref{fig6} the log-periodic interpretation of the oil price dynamics over the time period 2000-2010. This figure was publicly disclosed as an insertion to ref.~\cite{Stan08} and is shown here unchanged . We have two time scales here: A longer one ( red curve) with $T_c$=19.Sep.2010 and a shorter one (green curve) with $T_c$=11.Jul.2008. The green curve starts in the second minimum of the red curve and describes the sharp increase of the oil price since summer 2007. This green curve shows a clear log-periodic structure with our preferred $\lambda=2$, indicating the end of this increase of the oil market. Such a behavior is called \emph{Super-Bubble} and was first discussed in \cite{Stan03} in connection with a similar structure in the Nasdaq composite index. We typically do not reach such a perfect agreement with the actual market. But in the present case our $T_c$ was exactly the day, where the oil price approached its maximum and started to decline sharply. 
\section{Summary}

We have investigated some specific features of two complex systems. In the nuclear many-body problem we showed that the well know giant resonances that exist in all medium and heavy mass nuclei may be interpreted as a coherent structure embedded in chaos. The coherent part is generated by long range correlations, which appear mathematically by diagonalizing a Hamiltonian in a 1p1h space. 

In terms
of eigenvalues the collective state manifests itself through the gap relative to the remaining 
eigenvalues. Analog to such calculations in nuclei we investigated correlations in the dynamics of financial evolution. We constructed and diagonalized the correlation matrix for the DAX and Dow Jones indices and obtained results very similar to coherent states in nuclear physics. Also here the collective market component
is associated with the eigenvalue whose magnitude is visibly separated from the other more 'noisy'
eigenvalues. Even more, this separation is more pronounced during the market drawdowns than during
drawups which is consistent with the well known fact that the drawdowns are more collective
in the sense of coherence among the individual stocks forming the global market. 
We also saw that by properly taking into account the time-zone delay the both markets considered largely merge into a single one. 
Finally by discussing the phenomenon of financial log-periodicity. We indicated the criticality
characteristics of the financial dynamics and pointed to a predictive potential of the corresponding 
methodology.   
\section{Acknowledgments} 
J.S. thanks the Foundation for Polish Science for financial support through the Alexander von Humbold Honorary Research Fellowship.

\end{document}